\documentstyle[aps,epsf]{revtex}

\begin{document}
\title{Recent Results from Fermilab Charm Experiment E791}
\author{David A. Sanders}
\address{University of Mississippi}
\address{Representing the Fermilab Experiment E791 Collaboration}
\maketitle

\begin{abstract}
We report the results of some recent E791 charm analyses. They include 1) a
search for rare and forbidden decays, 2) measurements of form factors for 
$D^{+}\rightarrow K^{*0}\ell^{+}\nu _{\!_{\ell}}$ and 
$D_{\hbox{s}}^{+}\rightarrow \phi \ell^{+}\nu _{\!_{\ell}}$, and 3) 
$D_{\hbox{s}}^{+}$ and $D^{0}$ lifetime measurements including the lifetime 
difference between $D^{0}\rightarrow K^{-}\pi ^{+}$ and 
$D^{0}\rightarrow K^{-}K^{+}$. The latter is the first direct search for a 
possible lifetime difference that could contribute to 
$D^{0}-\overline{D}^{\,0} $ mixing. \ 
\end{abstract}


\baselineskip 14pt

%


\section{Introduction}

E791 is a high statistics charm experiment that acquired data at
Fermilab during the 1991-1992 fixed-target run. The experiment combined a
fast data acquisition system with an open trigger. Over $2\times 10^{10}\ $
events were collected with the Tagged Photon Spectrometer\cite{e791spect} using 
a 500 GeV $\pi ^{-}$ beam. There were five target foils with 15 mm 
center-to-center separations: one 0.5 mm thick platinum foil followed by 
four 1.6 mm thick diamond foils. The spectrometer included 23 planes of silicon 
microstrip detectors (6 upstream and 17 downstream of the target), 2 dipole 
magnets, 10 planes of proportional wire chambers (8 upstream and 2 downstream of 
the target), 35 drift chamber planes, 2 multi-cell \v {C}erenkov counters that
provided $\pi /K$ separation in the 6-60 GeV/$c$ momentum range \cite{Bartlett}, 
electromagnetic and hadronic calorimeters, and a muon detector.

\section{Rare and Forbidden Decays of $D^{+}$, $D_{s}$ and $D^{0}$}

We report the preliminary results of a search for flavor-changing 
neutral-current (FCNC), lepton number violating (LNV) and lepton family violating 
(LFV) decays of $D^+$, $D_{\hbox{s}}^{+}$ and $D^0$, into modes\footnote{all references to $D^+$, 
$D_{\hbox{s}}^{+}$ and $D^0$ and their decay modes also imply the corresponding 
charge-conjugate states.} containing muons and electrons. 
This analysis is an extension of our previous work\cite{FCNC}. The 
search for rare and forbidden decays provides another test of the Standard Model. 
The contribution, within the Standard Model, of short distance electroweak 
processes to the branching ratios of the FCNC-like decay modes described here 
is expected\cite{Schwartz} to be less than $10^{-8}$. The LNV and LFV 
decay modes are totally forbidden by the Standard Model. Therefore, the search 
for these rare and forbidden decay modes allows tests for violations of the 
Standard Model from FCNC, neutrino oscillations or other even more exotic 
processes. We are currently examining all the modes described in 
Table \ref{modes}; however, we are only presenting the following searches in 
this paper: $D^{+}\rightarrow \pi ^{+}e^{+}e^{-}$, $D^{+}\rightarrow \pi ^{+}\mu 
^{\pm }e^{\mp }$, $D^{+}\rightarrow \pi ^{-}e^{+}e^{+}$, $D^{+}\rightarrow 
\pi ^{-}\mu ^{+}e^{+}$, $D^{0}\rightarrow e^{+}e^{-}$, and $D^{0}\rightarrow 
\mu ^{\pm }e^{\mp }$. 

 \begin{table}
 \label{modes}
 \caption{Rare and Forbidden Decay Modes.}
 \renewcommand{\arraystretch}{1.15}
 \begin{tabular}{lll} 
 FCNC&LFV&LNV\\ 
 \tableline 
 $D^{+}\rightarrow \pi ^{+}\mu ^{+}\mu ^{-}$&
 $D^{+}\rightarrow \pi ^{+}\mu ^{\pm }e^{\mp }$&$D^{+}\rightarrow \pi ^{-}\mu ^{+}\mu ^{+}$\\
 $D^{+}\rightarrow \pi ^{+}e^{+}e^{-}$&$D^{+}\rightarrow \pi ^{-}\mu ^{+}e^{+}$
 &$D^{+}\rightarrow \pi ^{-}e^{+}e^{+}$\\
 $D_{\hbox{s}}^{+}\rightarrow K^{+}\mu ^{+}\mu ^{-}$&$D_{\hbox{s}}^{+}\rightarrow K^{+}\mu 
 ^{\pm }e^{\mp }$&$D_{\hbox{s}}^{+}\rightarrow K^{-}\mu 
 ^{+}\mu ^{+}$\\
 $D_{\hbox{s}}^{+}\rightarrow K^{+}e^{+}e^{-}$&$D_{\hbox{s}}^{+}\rightarrow K^{-}\mu 
 ^{+}e^{+}$&$D_{\hbox{s}}^{+}\rightarrow K^{-}e^{+}e^{+}$\\
 $D^{0}\rightarrow \mu ^{+}\mu ^{-}$&$D^{0}\rightarrow \mu ^{\pm }e^{\mp }$&\\
 $D^{0}\rightarrow e^{+}e^{-}$&&\\
 \end{tabular}
 \end{table}

To separate charm candidates from background, we require the secondary vertex
to be well-separated from the primary vertex and located well outside the
target foils and other solid material, the momentum vector of the candidate
$D^+$ or $D^{0}$ to point back to the primary vertex, and the decay track 
candidates to pass closer to the secondary vertex than the primary vertex. 
Specifically, the secondary vertex must be separated by $ > 20\,\sigma_L$ 
for $D^+$, ($ > 12\,\sigma_L$ for $D^{0}$) from the primary vertex and by 
$ > 5.0\,\sigma_L$ from the closest material in the
target foils, where $\sigma_L$ in each case is the calculated longitudinal
resolution in the measured separation.
The sum of the momentum vectors of the three tracks from this secondary
vertex must miss the primary vertex by $ < 40~\mu$m in the plane
perpendicular to the beam. 
We form the ratio of each track's smallest distance from the secondary vertex
to its smallest distance from the primary vertex, and require the product of
these ratios to be less than $10^{-n}$ where $n$ is the number of 
tracks in the secondary vertex. There is also a cut on the lifetime of the 
$D^+$ or $D^{0}$ candidate; it is $ < 5$ picoseconds for the $D^+$ and 
$ < 3$ picoseconds for the $D^0$. Finally, the net momentum of the $D^+$ 
candidate transverse to the line connecting the primary and secondary vertices 
must be less than 250 MeV/$c$ (300 MeV/$c$ for the $D^{0}$).

For this analysis we use a ``blind'' analysis technique. This method 
entailed the following steps: first covering the signal region with a ``box'', 
then optimizing ALL cuts using Monte Carlo simulated signal events and 
data from the wings outside the ``box'' before opening the ``box'' and only then 
opening the ``box'' that covers the signal region. The closed ``box'' 
mass windows are: $1.84<M(D^{+})<1.90$ GeV/$c^{2}$ for  $D^{+}\rightarrow 
\pi \mu \mu $, $1.78<M(D^{+})<1.90$ GeV/$c^{2}$ for  $D^{+}\rightarrow 
\pi ee$ and $\pi \mu e$, $1.95<M(D_{\hbox{s}}^{+})<1.99$ GeV/$c^{2}$ for 
$D_{\hbox{s}}^{+}\rightarrow K\mu \mu$, $1.91<M(D_{\hbox{s}}^{+})<1.99$ GeV/$c^{2}$ for 
$D_{\hbox{s}}^{+}\rightarrow Kee$ and $K\mu e$, $1.83<M(D^{0})<1.90$ GeV/$c^{2}$ for 
$D^{0}\rightarrow \mu \mu$, and $1.76<M(D^{0})<1.90$ GeV/$c^{2}$ for  
$D^{0}\rightarrow ee$ and $\mu e$. One should note the asymmetric windows for 
the decay modes containing electrons. This is to account for the electron 
bremsstrahlung tail. The upper limit of the branching ratio is 
calculated using the following formulae:
\begin{eqnarray}
\label{BReqn}
BR_{X}=\frac{N_{X}/\varepsilon _{X}}{N_{norm}/\varepsilon _{norm}}
\cdot BR_{norm}=\frac{N_{X}}{N_{norm}}\frac{\varepsilon _{norm}}
{\varepsilon _{X}}\cdot BR_{norm}
\end{eqnarray}
where $N_{X}$ is the number of observed events for a given decay 
mode and $\varepsilon $ is the 
detection efficiency. The ratio of the detector efficiencies for 
the normalization decay mode versus the unknown decay mode is
\begin{eqnarray}
\label{Effyeqn}
\frac{\varepsilon _{norm}}{\varepsilon _{X}}=\frac{N_{norm}^{MC}}{N_{X}^{MC}}.
\end{eqnarray}
$N_{X}$ is corrected, using the method of Feldman and Cousins\cite{Cousins}, 
for background. We have not yet corrected for systematic errors. The number of 
normalization events for the $D^{+}$ decay modes is 
$N_{norm}=25330\pm 169$, $D^{+}\rightarrow K^{-}\pi ^{+}\pi ^{+}$ 
events, and for the $D^{0}$ decay modes is $N_{norm}=26980\pm 185$, 
$D^{0}\rightarrow K^{-}\pi ^{+}$ events.

\begin{figure}[h]	
\centerline{\epsfxsize 7.1 truein \epsfbox{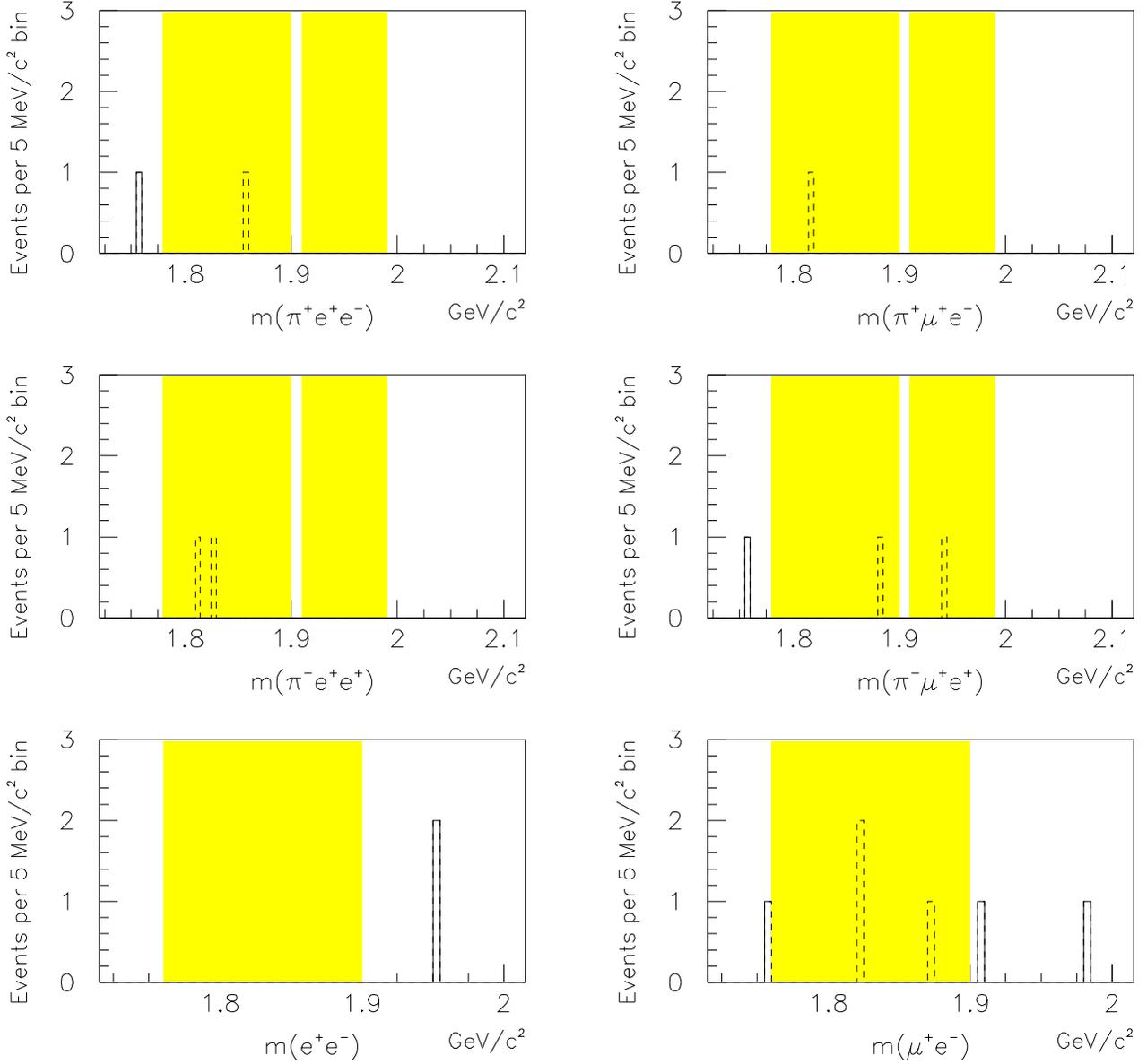}}   
\vskip -.2 cm
\caption[]{
\label{Data}
\small The preliminary data for the $D^+$ and $D^0$ decay modes presented here.}
\end{figure}

In addition to the cuts made on the kinematic variables described 
above, particle identification cuts are used to reduce the background. 
These cuts tagged the dilepton candidates, thereby greatly reducing the 
background.  Events which could also be reconstructed as a reflection, 
due to particle misidentification, from other purely hadronic decay modes 
into a region outside the ``boxes'', but within the overall mass 
window, were removed. Examples of this would include 
$D^{+}\rightarrow K^{-}\pi ^{+}\pi ^{+}$ reconstructed as $D^{+}\rightarrow 
\pi ^{-}\ell^{+}\ell^{+}$ or $D^{0}\rightarrow K^{-}\pi ^{+}$ reconstructed as 
$D^{0}\rightarrow \ell^{+}\ell^{-}$. In Figure \ref{Data}, we present 
preliminary plots of the data for the $D^+$ and $D^0$ decay modes 
described in this paper. The yellow boxes are the part of the plot 
that were originally covered by the ``boxes''. The events observed 
within these ``boxes'' are the total number of observed events, signal 
plus background. In the plots of the $D^{+}$ decay modes the lower 
yellow box is the $D^{+}$ ``box'' and the upper yellow box is the 
$D_{\hbox{s}}^{+}$ ``box''.

We estimate the mean background assuming a flat distribution calculated 
from the events seen in the regions that are outside the ``boxes''. 
Then the Poisson signal mean $\mu $ for the total number of observed events for 
a 90$\%$ confidence level upper limit are calculated, using the method of Feldman and 
Cousins\cite{Cousins}, and this number, $N_{X}$, is substituted into Equation 
\ref{BReqn} thus giving a 90$\%$ confidence level upper limit branching 
ratio. The estimated background, total number of events within the ``box'', 
the number $N_{X}$ for a 90$\%$ confidence level upper limit, the calculated 
upper limit branching ratios and the previously measured upper limits\cite{PDG} 
are given in Table \ref{Results}.

\vfill
\eject

\begin{figure}[h]	
\centerline{\epsfxsize 5.5 truein \epsfbox{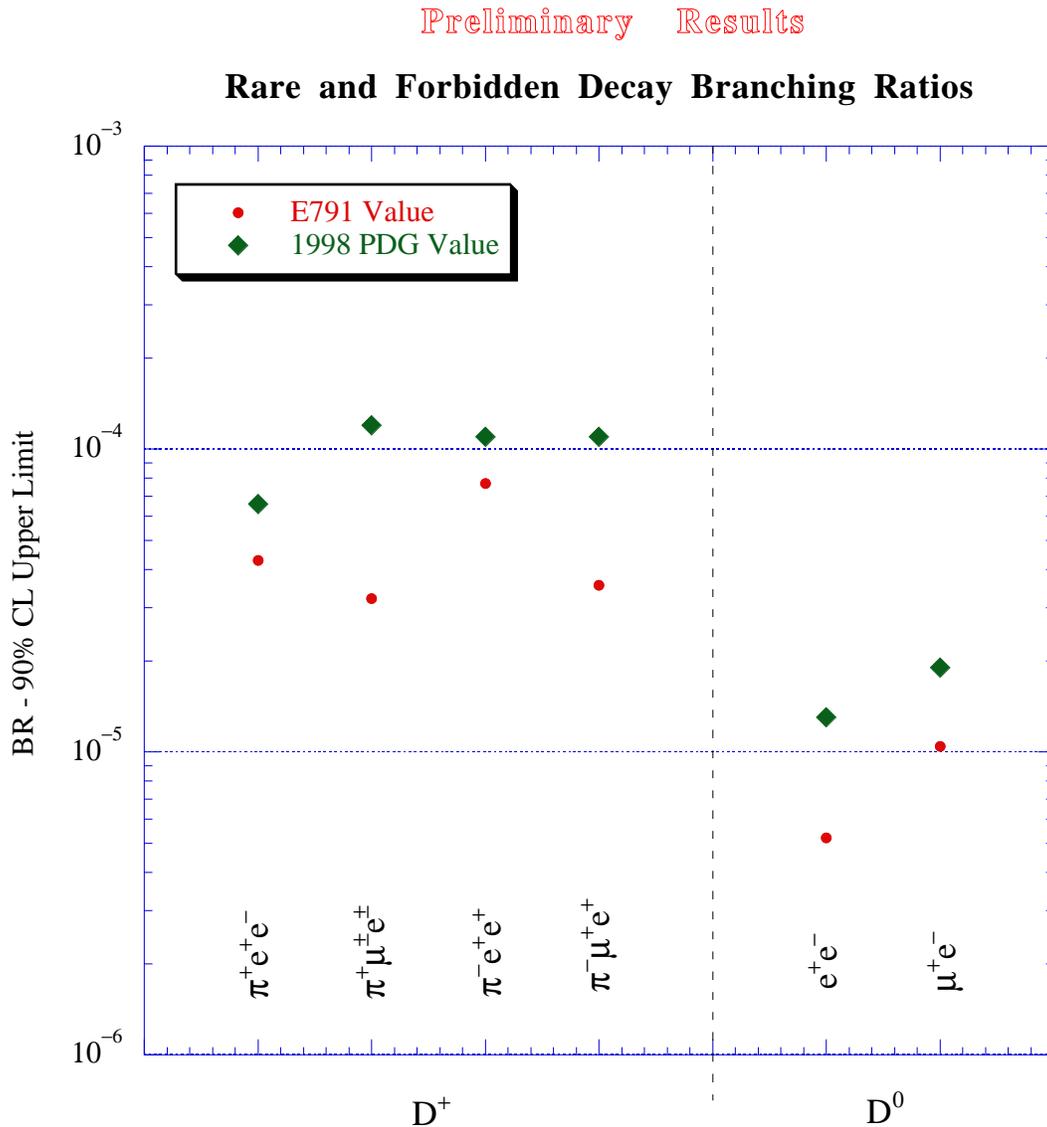}}   
\vskip -.2 cm
\caption[]{
\label{BR}
\small The 90$\%$ C.L. upper limit branching ratios for E791 data and for 
the 1998 PDG values.}
\end{figure}

 \begin{table}
 \label{Results}
 \caption{Preliminary Results Ð 90$\%$ C.L. Upper Limits}
 \renewcommand{\arraystretch}{1.15}
 \begin{tabular}{ldcdll} 
 Mode&Estimated Background
 &Observed&Calculated\protect \tablenote{By Method of Feldman and 
 Cousins. (90$\%$ C.L.)} 
 $N_{X}$&Branching Ratio&1998 PDG BR\\
 \tableline
 $D^{+}\rightarrow \pi ^{+}e^{+}e^{-}$&0.6&1&3.76&$<4.3\times 10^{-5}$
 &$<6.6\times 10^{-5}$\\
 $D^{+}\rightarrow \pi ^{+}\mu ^{\pm }e^{\mp }$&0.0&1&4.36&$<3.2\times 10^{-5}$
 &$<1.2\times 10^{-4}$\\
 $D^{+}\rightarrow \pi ^{-}e^{+}e^{+}$&0.0&2&5.91&$<7.7\times 10^{-5}$
 &$<1.1\times 10^{-4}$\\
 $D^{+}\rightarrow \pi ^{-}\mu ^{+}e^{+}$&0.6&1&3.76&$<3.6\times 10^{-5}$
 &$<1.1\times 10^{-4}$\\
 \tableline 
 $D^{0}\rightarrow e^{+}e^{-}$&1.8&0&1.30&$<5.2\times 10^{-6}$
 &$<1.3\times 10^{-5}$\\
 $D^{0}\rightarrow \mu ^{\pm }e^{\mp }$&2.6&3&4.80&$<1.0\times 10^{-5}$
 &$<1.9\times 10^{-5}$\\
 \end{tabular}
 \end{table}
A comparison between our calculated 90$\%$ confidence 
level upper limit branching ratios and the previous results\cite{PDG} 
is shown by the plot of the 90$\%$ C.L. upper limits in Figure \ref{BR}.

\vfill
\eject

\section{$D_{s}$ Lifetime Measurements}

Accurate measurements of the lifetimes of the weakly decaying charm mesons
are useful for understanding the contributions of various weak decay
mechanisms. Despite the fact that they are all tied to the charm quark
decay, the decays of the ground state charm mesons can have 
$\em{different}$ contributions from the four first-order processes (two spectator,
W-annihilation, and W-exchange) and their lifetimes are, in fact, quite
varied\cite{PDG}.

\begin{eqnarray}
\tau \left( D^{+}\right) :\tau \left( D^{0}\right) :\tau \left( D_{\hbox{s}}\right)
=2.5:1:1.1
\end{eqnarray}

We have made a new precise measurement of the $D_{\hbox{s}}$ lifetime\cite{DsLife}
using $1662\pm 56$ fully reconstructed $D_{\hbox{s}}^{+}\rightarrow \phi \pi ^{+}$ decays.
The $D_{\hbox{s}}$ lifetime is measured using an unbinned maximum-likelihood fit to
be $0.518\pm 0.014\pm 0.007$ ps. This value is somewhat higher than the
world average $D^{0}$ lifetime \cite{PDG} of $0.467\pm 0.017$ ps. Using our
result and the world average $D^{0}$ lifetime \cite{PDG}, we find that the
ratio of the $D_{\hbox{s}}$ lifetime to the $D^{0}$ lifetime to be

\begin{eqnarray}
\frac{\tau \left( D_{\hbox{s}}\right) }{\tau \left( D^{0}\right) }=1.25\pm 0.04%
\text{ (a 6}\sigma \text{ difference from unity)}
\end{eqnarray}
showing, for the first time, significantly different lifetimes for the 
$D_{\hbox{s}}$ and $D^{0}$. This result may be used to constrain the 
contributions of various decay mechanisms to charm decay and further refine our 
quantitative understanding of the hierarchy of charm particle lifetimes.

\section{Measurement of the Lifetime Difference Between $D^{0}\rightarrow
K^{-}\pi ^{+}$ and $D^{0}\rightarrow K^{-}K^{+}$}

We report the first directly measured constraint on the decay-width
difference $\Delta \Gamma $ for the mass eigenstates of the 
$D^{0}-\overline{D}^{\,0}$ system \cite{D0mix}. We obtain our result from 
lifetime measurements of the decays $D^{0}\rightarrow K^{-}\pi ^{+}$ and 
$D^{0}\rightarrow K^{-}K^{+}$, under the assumption of $CP$ invariance, which 
implies that the $CP$ eigenstates and the mass eigenstates are the same. The 
lifetime of the $CP$-even final state $D^{0}\rightarrow K^{-}K^{+}$, as 
calculated from $6683\pm 161$ weighted events, is 
$\tau _{KK}=0.410\pm 0.011\pm 0.006$ ps. The lifetime of an equal mixture of 
$CP$-odd and $CP$-even final states $D^{0}\rightarrow K^{-}\pi ^{+}$, as 
calculated from $60581\pm 353$ weighted events, is 
$\tau _{K\pi }=0.413\pm 0.003\pm 0.004$ ps. We find that 
$\Delta \Gamma =2\left( \Gamma _{KK}-\Gamma _{K\pi }\right) $ 
$=0.04\pm 0.14\pm 0.05$ ps$^{-1}$, leading to a limit of 
$-0.20<\Delta \Gamma <0.28$ ps$^{-1}$ at 90\% C.L. The value of $\Delta \Gamma $ 
is consistent with zero and thus, at our level of sensitivity, is consistent with 
the Standard Model.

\section{Measurements of $D^{+}\rightarrow K^{*0}\ell^{+}\nu _{\!_{\ell}}$ and 
$D_{s}^{+}\rightarrow \phi \ell^{+}\nu _{\!_{\ell}}$ Form Factors}

The Vector and Axial form factors are used to extract elements of the
Cabibbo-Kobayashi-Maskawa mixing matrix. Here we present measurements of the
form factor ratios, evaluated at $q^{2}=0$ $($GeV$/c)^{2}$, $r_{V}=V\left(
0\right) /A_{1}\left( 0\right) $, $r_{2}=A_{2}\left( 0\right) /A_{1}\left(
0\right) $, for both the muon and electron channels, and 
$r3=A_{3}\left( 0\right) /A_{1}\left( 0\right) $ in the muon channel. This is 
the first simultaneous measurement of both the muon and electron channels.

\subsection{$D^{+}\rightarrow K^{*0}\ell^{+}\nu _{\!_{\ell}}$ Form Factor Ratios}

We have already reported the form factor ratios $r_{V}$ and $r_{2}$ for the
decay $D^{+}\rightarrow K^{*0}e^{+}\nu _{e}$, $K^{*0}\rightarrow K^{-}\pi %
^{+}$ \cite{FormFactD+e} . The form factor ratios in the decay $%
D^{+}\rightarrow K^{*0}\mu ^{+}\nu _{\mu }$, $K^{*0}\rightarrow K^{-}\pi ^{+}
$  were measured using $3034(595)$ signal (background) events and yield\cite
{FormFactD+} $r_{V}=1.84\pm 0.11\pm 0.09$, $r_{2}=0.75\pm 0.08\pm 0.09$, and
as a first measurement $r_{3}=0.04\pm 0.33\pm 0.29$. The combined electron
and muon form factor ratios, for approximately 6000 events, are $r_{V}=1.87%
\pm 0.08\pm 0.07$ and  $r_{2}=0.73\pm 0.06\pm 0.08$.

\subsection{$D_{s}^{+}\rightarrow \phi \ell^{+}\nu _{\!_{\ell}}$ 
Form Factor Ratios}

We have measured the form factor ratios $r_{V}$ and $r_{2}$ for the decay 
$D_{\hbox{s}}^{+}\rightarrow \phi \ell^{+}\nu _{\!_{\ell}}$, $\phi \rightarrow K^{-}K^{+}$ 
in both the electron and muon channels\cite{FormFactDs}. The form factor ratios 
in the electron channel were measured using $144(22)$ signal (background)
events and yield $r_{V}=2.24\pm 0.47\pm 0.21$ and $r_{2}=1.64\pm 0.34\pm 0.20
$. In the muon channel there were $127(34)$ signal (background) events,
yielding $r_{V}=2.32\pm 0.54\pm 0.26$ and $r_{2}=1.49\pm 0.36\pm 0.20$. The
combined electron and muon form factor ratios, for approximately 271 events,
are $r_{V}=2.27\pm 0.35\pm 0.22$ and $r_{2}=1.57\pm 0.25\pm 0.19$.

\bigskip 

We gratefully acknowledge the assistance of the staffs of Fermilab and
of all the participating institutions. This research was supported by
the Brazilian Conselho Nacional de Desenvolvimento Cient\'\i fico e
Tecnol\'ogico, CONACyT (Mexico), the Israeli Academy of Sciences and
Humanities, the U.S. Department of Energy, the U.S.-Israel Binational
Science Foundation, and the U.S. National Science Foundation. Fermilab
is operated by the Universities Research Association, Inc., under
contract with the United States Department of Energy.


\smallskip 

\begin{references}
\bibitem{e791spect}
 J. A. Appel, Ann. Rev. Nucl. Part. Sci. {\bf 42} 367-399 (1992), and references therein;\hfil\break
 D. J. Summers {\em et al.}, Proceedings of the {\it XXVII$^{\,th}$ Rencontre de  
   Moriond}, Electroweak Interactions and Unified Theories, Les Arcs, 
   France 417-422 (15-22 March, 1992);\hfil\break
 S. Amato {\em et al.}, Nucl. Instr. Meth. {\bf A324} 535-542 (1993);\hfil\break
 S. Bracker  {\em et al.}, IEEE Trans. Nucl. Sci. {\bf NS-43} 2457-2464 (1996);\hfil\break 
 E.M. Aitala {\em{et al.}} (E791), FERMILAB-PUB-98-297-E and
   hep-ex/9809029, submitted to Eur. Phys. J. {\bf C}.

\bibitem{Bartlett}  D. Bartlett {\em{et al.}}, Nucl. Instr. Meth. {\bf A260,} 
55-121 (1987).

\bibitem{FCNC}  E.M. Aitala {\em{et al.}} (E791), (hep-ex/9507003) Phys. Rev. Lett. {\bf 76,}
364-367 (1996).

\bibitem{Schwartz}  A.J. Schwartz, Mod. Phys. Lett. A {\bf 8,} 967-977 
(1993).

\bibitem{Cousins}  Gary J. Feldman and Robert D. Cousins, Phys. Rev. {\bf D57,} 
3873-3889 (1998).

\bibitem{PDG}  C. Caso et al. (Particle Data Group), Eur. Phys. J. {\bf C3,}
1 (1998).

\bibitem{DsLife}  E.M. Aitala {\em{et al.}} (E791), (hep-ex/9811016) 
Phys. Lett. {\bf B445,} 449-454 (1999).

\bibitem{D0mix}  E.M. Aitala {\em{et al.}} (E791), 
FERMILAB-PUB-99/036-E and hep-ex/9903012, submitted to Phys. Rev. Lett.

\bibitem{FormFactD+e}  E.M. Aitala {\em{et al.}} (E791), (hep-ex/9710216) 
Phys. Rev. Lett. {\bf 80,} 1393-1397 (1998).

\bibitem{FormFactD+}  E.M. Aitala {\em{et al.}} (E791), (hep-ex/9709026) 
Phys. Lett. {\bf  B440,} 435-441 (1998).

\bibitem{FormFactDs}  E.M. Aitala {\em{et al.}} (E791), (hep-ex/9812013) 
Phys. Lett. {\bf B450,} 294-300 (1999).
\end{references}
\end{document}